\documentclass[12pt]{iopart}
\usepackage{iopams}  
\usepackage{natbib}
\usepackage{graphicx}
\usepackage{aas_macros}

\begin{document}

\title[]{The Galactic Gravitational wave foreground}

\author{Gijs Nelemans}

\address{Department of Astrophysics, IMAPP, Radboud University
  Nijmegen\\PO Box 9010, 6500 GL, Nijmegen, the Netherlands}
\ead{nelemans@astro.ru.nl}
\begin{abstract}
  I present an overview of the Galactic binaries that form the
  foreground for the ESA/NASA Laser Interferometer Space Antenna
  (LISA). The currently known population is discussed, as well as
  current and near-future large-scale surveys that will find new
  systems. The astrophysics that can be done when the LISA data
  becomes available is presented, with particular attention to
  verification binaries, the overall Galactic populations, neutron
  star and black hole binaries and sources in globular clusters. I
  discuss the synergy with electro-magnetic observations and correct
  an error in the estimate of the number of LISA systems that can be
  found in the optical compared to \citet{nel06b} and conclude that
  at least several hundreds of systems should be detectable.
\end{abstract}


\submitto{\CQG}
\maketitle

\section{Introduction: Galactic gravitational wave binaries}

Already in the 1980s, it was realised that compact binaries in the
Galaxy are sources of low-frequency gravitational waves
\citep[e.g.][]{eis87,lpp87,hbw90}. In particular double white dwarf
binaries dominate the signal at frequencies above 0.1 mHz, with a
total Galactic population of $\sim$30 million objects \citep{hbw90,nyp03,trc05}.

The binaries that are most relevant for LISA are the so-called
ultra-compact binaries \citep{nel06}, consisting of two compact
evolved stars. We distinguish 
\begin{enumerate}
\item \emph{Detached} binaries where both stars are well separated, of
  which the following classes are known
\begin{description}
\item[Double white dwarfs] are the most common systems. They are the
  endpoints of many binary evolution scenario's
  \citep[see][]{web84}. They have been discovered observationally
  \citep[e.g.][]{slo88,mar95} and the largest project to find them,
  the ESO Supernova Ia Progenitor SurveY \citep[SPY;][]{ncd+01} has
  discovered several tens of them, although very few in the LISA
  frequency range \citep[see][]{nnk+05}.
\item[White dwarf -- neutron star binaries] are typically found as
  white dwarfs around radio pulsars
  \citep[see][]{2005LRR.....8....7L}, with long-period orbits. No
  systems in the LISA frequency range are known, but several are
  expected \citep{nyp01,nel03b}.
\item[Double neutron stars] were the first to be discovered
  \citep{ht75} but currently only 8 are known
  \citep[see][]{2005LRR.....8....7L}, of which the shortest period is
  2.4 hours.
\end{description}
\item \emph{Interacting} binaries, in which mass is transferred from
  one star to the other, of which the following classes are known
\begin{description}
\item[AM CVn stars,] in which a white dwarf is accreting (helium rich)
  material from a compact companion \citep[e.g.][]{war95,nel05}. Currently 22
  are known with periods between 5.4 and 65 min \citep[see][]{nel05}.
\item[Ultra-compact X-ray binaries,] which have neutron star accretors
  (no ultra-compact black hole systems have yet been found). Currently
  there are 27 (candidate) systems \citep{zjm+07}, only 8 with well
  known periods between 11 and 50 minutes. From their X-ray and
  optical spectra it has been inferred that the donors stars can be
  either helium rich or carbon/oxygen rich
  \citep{sch+01,jpc00,njm+04,njs06} but from the properties of type I
  X-ray bursts it has been inferred that in most systems the matter
  accumulated on the neutron star is helium \citep[see][]{icv05}.
\end{description}
\end{enumerate}

Ultra-compact binaries, especially when observed with LISA, are interesting
objects from Astrophysical point of view in a number of areas.
\begin{description}
\item[Binary star evolution] The ultra-compact binaries represent one of the
  most evolved stages in binary evolution and in order to get their
  ultra-short periods, they must have had extreme angular momentum loss in one
  or likely more common-envelope phases \citep[e.g.][]{nrs81,web84}. The
  process of angular momentum loss in a common envelope is poorly understood
  \citep[e.g.][]{ts00b,nt03} so detailed understanding of ultra-compact
  binaries will teach us something about binary evolution in general. At the
  short periods that these binaries have, angular momentum loss due to
  gravitational wave radiation becomes the most important driver of the binary
  evolution \citep[e.g.][]{pac67,vil71,ty79a}.
\item[Type Ia Supernovae progenitors] The use of type Ia supernovae as
  standard candles \citep{phi93} has led to the discovery of the
  accelerated expansion of the Universe
  \citep{1998AJ....116.1009R,1999ApJ...517..565P}, but their
  progenitors are unknown. Directly observing the progenitors is
  difficult, as they are typically observable only nearby, where the
  occurrence rates are low \citep[of the order of a few per millennium
  in our Galaxy, e.g.][]{2008A&A...479...49B,2006ApJ...648..868S},
  except using archival Chandra X-ray or Hubble Space Telescope
  optical observations \citep{vn08,2008arXiv0801.2898M,rbv+08,nvr+08}. Therefore, studying
  the potential progenitor populations and determining their
  occurrence rates is a promising way forward
  \citep[e.g.][]{2005ASSL..332..163Y,2006MNRAS.368.1893F}. As such,
  ultra-compact binaries are very relevant, as some of them may be
  progenitors \citep[e.g.][]{2005ASPC..334..387S}, but they certainly
  come from the same general population in which the supernovae
  originate.
\item[Galactic structure] When LISA will discover many thousands of
  ultra-compact binaries \citep[e.g.][see Sect.~\ref{foregrounds}]{nyp01}, it
  opens up the possibility to chart their distribution throughout the Galaxy,
  in particular in the inner region, where most systems are expected. Thus,
  LISA measurements will contribute towards our understanding of the structure
  of the Galaxy. 
\item[Binary interaction] Finally, during the evolution of ultra-compact
  binaries, there may be other processes except for gravitational wave
  radiation and mass transfer that determine the orbital evolution, in
  particular tidal dissipation in the white dwarfs
  \citep[e.g.][]{rpa06,2008PhRvL.100d1102W}. LISA observations will thus allow
  detailed study of the physics of mass transfer, tides and other interactions
  in these ultra-compact binaries.
\end{description}

\section{Recent progress in Astrophysics: Surveys}

One of the recent developments in astrophysics is the advent of digital
large-scale surveys. 

\begin{description}
\item[SDSS] The  Sloan Digital Sky Survey \citep[{\sl
    SDSS/SEGUE}][]{2000AJ....120.1579Y}, is a $\sim$8\,000 square degree survey, which was largely aimed at
  extra-galactic sources. The area is imaged in five optical bands (u,g,r,i,z)
  yielding more than 250 million objects and of a subset of about 1 million
  sources spectra were taken. From the SDSS database 7 AM CVn systems were found
  \citep{rgm+05,ahh+05,2008AJ....135.2108A}. Extrapolating this to the full
  photometric database, another 40 systems should be in the {\sl SDSS}
  area. Low-resolution identification spectroscopy of the candidates is
  currently underway \citep{2008arXiv0811.3974R}. Comparing these numbers with
  the predictions of population models \citep{nyp03} it was found that even
  pessimistic models likely overestimate the true number of AM CVn stars in
  the Galaxy \citep{rng07}.
\item[RATS and OmegaWhite] The Rapid Temporal Survey {\sl RaTS}
  \citep{2005MNRAS.360..314R} and {\sl
    OmegaWhite}\footnote{http://www.astro.ru.nl/omegawhite} are
  variability surveys that specifically target short period
  ($\lesssim$1 hour) variability by taking time series photometry of
  large areas on the sky (40 and 400 square degrees respectively) for
  periods of 2 hours with points each few minutes. {\sl RaTS} is
  currently underway on the Isaac Newton Telescope and the ESO 2.2m
  telescope. {\sl OmegaWhite} will start in 2009 using OmegaCam
  \citep{2006SPIE.6276E...9I} on the ESO VLT Survey Telescope. Many
  new, short period ultra-compact binaries will be discovered.
\begin{figure}
\centerline{\includegraphics[angle=-90,width=0.8\textwidth]{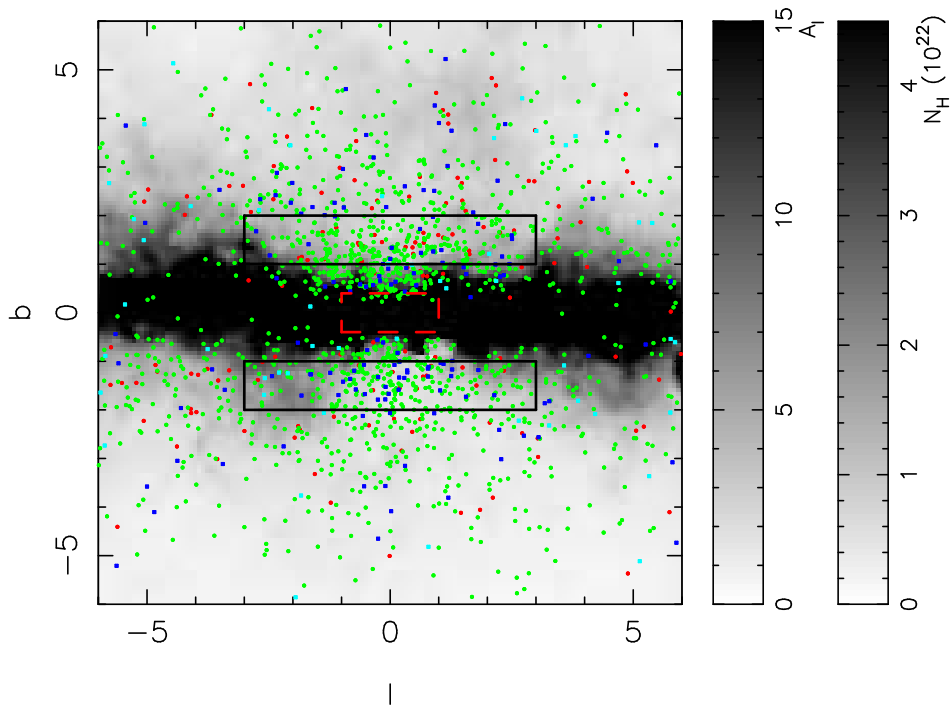}}
\caption[]{Area of the Galactic Bulge Survey in Galactic coordinates. The two
  boxes show the final area that will be observed, and have been chosen to
  avoid the largest extinction (shown as grey shade). The points
  indicate the population of expected X-ray sources, based on very
  simple estimates. The dashed box is the area of the \citet{wgl02}
  X-ray survey for which optical follow-up is impossible due to extinction.}
\label{fig:GBS}
\end{figure}
\item[GBS] The The Galactic Bulge survey ({\sl GBS}) uses the {\sl Chandra}
  X-ray satellite and ground-based optical observations to chart the
  population of X-ray binaries (including ultra-compacts) in the inner part of
  the Galactic Bulge (Jonker et al., in preparation). Observations are make of
  two regions of 1 $\times$ 6 square degrees each, located directly above and
  below the Galactic Center at 1$^\circ <|b|<$ 2$^\circ$ (Fig.~\ref{fig:GBS}).
  Using this observing strategy it avoids the very strong dust absorption in
  both X-rays as well as red optical wavelengths that plague observations at
  the very Galactic Center. The same area has been imaged using the CTIO 4-m
  Blanco telescope in the optical red broad-band $r,i$ and nar\-row-band
  H$\alpha$ filters to a depth of $r\sim$23.5 and equivalent H$\alpha$ line
  flux. This depth is chosen such that a low-mass main-sequence companion to a
  neutron star or black hole accretor at the distance of the Galactic Center
  will be detected. H$\alpha$ observations have been included since many
  non-ultra-compact binaries will show H$\alpha$ emission in their
  spectra. All optical observations have been obtained and are reduced. Half
  of the X-ray observations are obtained and reduced and already have led to
  the identification of 1\,700 new X-ray sources. Due to the exquisite spatial
  resolution of {\sl Chandra} many of these have unique counterparts in the
  optical data. The remaining X-ray observations will be obtained over the
  coming years. The number of X-ray sources is in line with our model
  calculations, giving confidence in the number of (ultra-)compact binaries to
  be detected from the {\sl GBS} (30-60).
\item[EGAPS] {\sl The European Galactic Plane Survey (EGAPS)} is surveying the
  full Galactic Plane in a strip of 10$\times$360 square degrees centered on
  the Galactic equator. It uses the broad-band optical $U,g,r,i$ bands and
  additionally the narrow-band H$\alpha$ and He{\sc i} 5875 (northern
  hemisphere only) bands, down to 21$^{st}$ magnitude or equivalent line
  flux. The survey has been described in \citet{2005MNRAS.362..753D} and Groot
  et al. (submitted). The Northern survey has been running on the Isaac
  Newton Telescope on La Palma since summer 2003 and is currently 65\%
  complete. The southern survey will start next year on the VST survey
  telescope of the European Southern Observatory as a 100-night ESO Public
  Survey.
\item[GAIA] The ESA GAIA mission \citep{2001A&A...369..339P} will
  image the whole sky down to magnitudes around V=20 with the aim to
  measure positions, spectral energy distributions and radial
  velocities (for the brighter stars) of upto one billion stars. The
  satellite will continuously map the sky while it rotates and thus
  build up an enormous set of very accurate relative position
  measurements. At the end these can be turned into absolute
  positions. As each position in the sky is visited many times
  (typically around 90 times) parallax and proper motion of all
  objects will be determined.
\end{description}

\section{Astrophysics with LISA}

\subsection{Verification binaries}

\begin{figure}
\centerline{\includegraphics[angle=-90,width=0.8\textwidth]{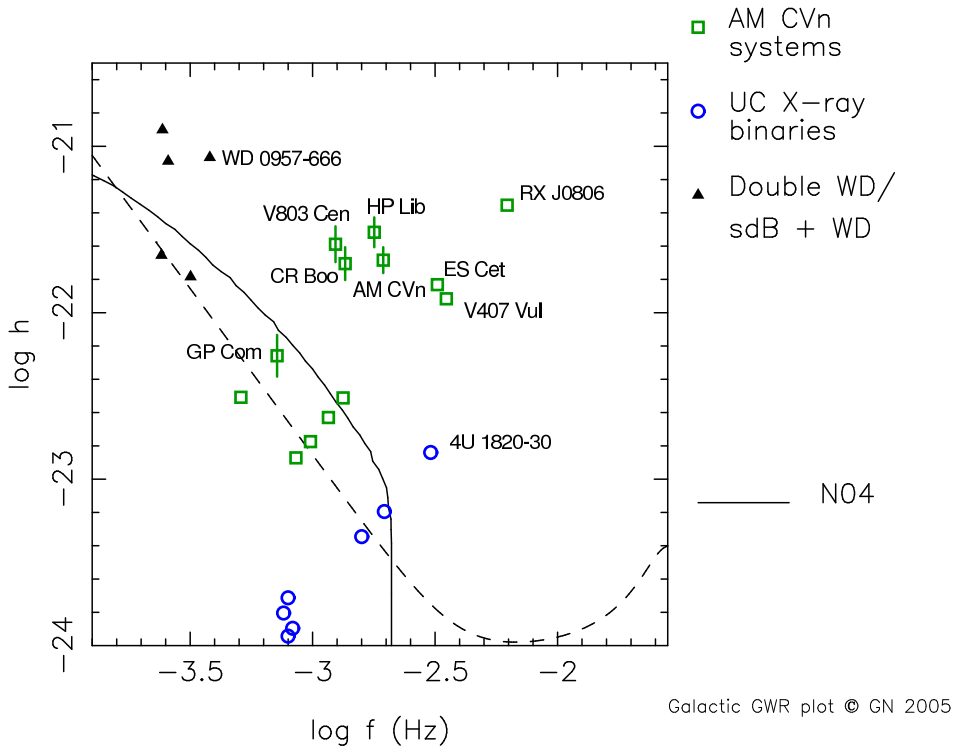}}
\caption[]{Expected LISA signal of several known ultra-compact
  binaries, as indicated. Only for a handful of AM CVn stars their
  parameters are well determined and reliable error bars can be given
  \citep{rgb+07}, the rest are estimates. The dashed line shows the
  LISA sensitivity, the solid line an estimate of the Galactic
  foreground noise \citep[from][]{nyp03}.}
\label{fig:verification}
\end{figure}

One of the uses of ultra-compact binaries is that some are
\emph{known, guaranteed} LISA sources and thus can be used as
verification sources \citep[e.g.][]{sv06} even though (much) stronger,
but yet unknown, sources will likely be detected first. Important for
the use of known sources as verification is of course to know their
properties as accurately as possible before LISA flies. Recent
progress here has been made by using the FGS instrument on board of
the Hubble Space Telescope to measure accurate distances to a number
of AM CVn stars \citep{rgb+07}. Together with estimates for the
component masses from the absolute magnitudes this has led to
estimates of the expected LISA signal that are well determined with
reliable error bars \citep[Fig.~\ref{fig:verification}
and][]{rgb+07}. For the shortest periods systems the distances and
component masses are not (yet) well determined enough to give well
defined signal estimates.

\subsection{(Un)resolved foregrounds}\label{foregrounds}

The number of ultra-compact binaries depends strongly on their orbital
period as the evolutionary timescale decreases sharply towards shorter
periods. At the same time the frequency resolution of LISA remains
constant. Therefore there is a large difference in the properties of
the ultra-compact binary population at low frequencies, where there
are many objects in the Galaxy per frequency resolution element, and
at high frequencies, where there are few, if any objects per
resolution element \citep[e.g.][]{eis87}.

At high frequencies many systems can in principle be individually detected and
have their properties measured with high accuracy, depending on the
signal-to-noise ratio. The current Mock LISA data challenges
\citep[e.g.][]{2007CQGra..24..529A} show that indeed in more or less realistic data
sets, the ultra-compact binaries can be detected. As these are a small subset
of the population and the total population is probably strongly concentrated
in the inner Galaxy, most of these systems will reside close to the Galactic
center.  For some fraction of these systems the frequency evolution can be
detected \citep[e.g.][]{svnw05}, although the details are still under
investigation in the MLDC rounds.

A separate class of systems that will be individually detected are lower
frequency systems that have such strong signals that they stick out above the
local noise that is formed by the collective weaker signal sources. These are
typically the intrinsically stronger sources (with high-mass components) and
the nearby sources \citep[e.g.][]{2007CQGra..24..513B}.

At lower frequencies the many systems together form what is often called the
unresolved Galactic foreground (a better name than the previously used
``background''). Although this is often depicted as an extra noise component
that at low frequencies exceeds the instrumental noise, this is a bit
misleading as it really is a signal and more importantly, it is variable over
the year as the bulk of this foreground comes from systems that are located
towards the Galactic center \citep[e.g.][]{etk+05}. This opens the possibility
to use its shape to learn about the distribution of the sources in the
Galaxy. In particular the different Galactic components (thin disk, thick
disk, halo) have a contribution that vary differently, although the
contribution of thick disk and halo are very small compared to the thin disk
(e.g. Ruiter et al. ApJ submitted).

\subsection{Electro-magnetic counterparts}

\begin{figure}
\centerline{\includegraphics[angle=-90,width=0.8\textwidth]{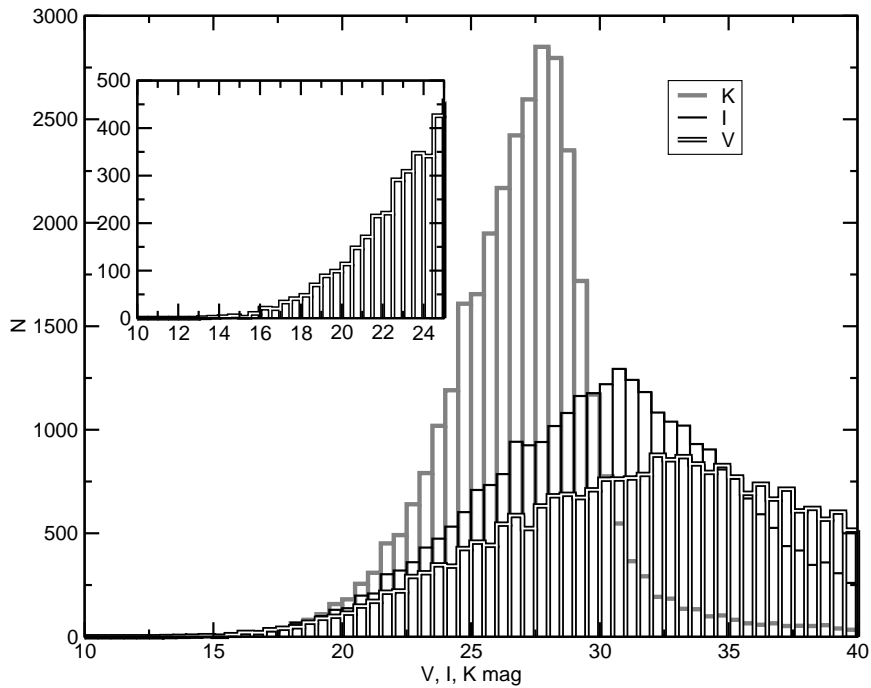}}
\caption[]{Histogram of apparent magnitudes of the double white dwarfs
  that are estimated to be individually detected by LISA (see
  text). The black-and-white line shows the distribution in the
  V-band, the black line the distribution in the I-band and the grey
  histogram the distribution in the K-band. Galactic absorption is
  taken into account. The zoom-in shows the bright-end tail of the
  V-band distribution.}
\label{fig:dwd_VIK}
\end{figure}

A very interesting possibility is to observe systems both individually
with LISA, as well as with electro-magnetic means. The information
that can be obtained from the gravitational-wave data is complementary
to that which can be obtained from electro-magnetic data, in
particular optical and/or near infrared. The question is how many of
the systems that LISA will detect are also observable with optical or
near infrared detectors. The problem is that most resolved binaries
reside close to the Galactic center, where they suffer from heavy
extinction. \citet{cfs03} estimated that a large fraction (several
tens of percents) of the double white dwarf LISA sources would be
detectable electro-magnetically, but we showed \citep{nel06b} that
they likely overestimated the intrinsic brightness of the white dwarfs
and that only several tens of systems would be detectable. However, we
recently found that that estimate is too pessimistic due to an error
in the Galactic distribution of the systems in \citet{nel06b}, which
is too concentrated to the Galactic center\footnote{Due to an
  underestimate of the formation time of the systems in the Galaxy for
  which we use a star formation history that starts in the inner
  Galaxy}. We redid the calculation and now find the much more
optimistic result that several hundreds of double white dwarf LISA
systems might be detectable electro-magnetically
(Fig.~\ref{fig:dwd_VIK}). Note that for this estimate we use a very
simplistic estimate of the number of systems that can be individually
observed with LISA: all systems with frequencies above 2.1mHz or
(barycentric) strain amplitudes larger with $\log h > -28.152 - 1.9992
\log f$. This results in a 33 thousand systems of which $\sim$2000
have V $<$ 24 ($\sim$6\%). This simple estimate for the number of
individually resolved systems is likely an overestimate
\citep[e.g.][]{2007CQGra..24..513B}. We are currently investigating
this in more detail (Finn \& Nelemans, in preparation). Our new
estimate is still lower than \citet{cfs03}, because of their
overestimate of the instrinsic brightness of white dwarfs.

In Fig.~\ref{fig:dwd_VIK} we also show the I-band and K-band magnitudes of the
LISA systems that suffer less from extinction. For apparent magnitudes fainter
than 20 the number of systems is largest in the K-band. Current wide field
surveys and instruments typically have limiting magnitudes of 22-24 in the V-
and I-bands and 20-21 in the K-band.

As also mentioned by \citet{cfs03} one of the most interesting features of the
double white dwarfs that can be detected electro-magnetically is that a large
fraction of them will be very short period, with relatively large probability
of showing eclipses. In this respect the GAIA measurements, which typically
consist of $\sim$90 photometric measurements are interesting, as
eclipsers should show up in this data (Marsh \& Nelemans, in
preparation). Detecting eclipsers would allow to get detailed information
about the absolute dimensions of the systems and might allow detection of
period evolution even for the systems for which the LISA mission duration is
too short to measure it.

\subsection{Neutron star and black hole binaries}

Not that much attention has been been given to ultra-compact binaries with
neutron star and black hole components for LISA. They are typically considered
for the ground based detectors as they are the only systems at the high
frequencies accessible from Earth. And although the numbers are much smaller
than those with white dwarf components, there are several tens of systems
expected \citep[e.g.][]{nyp01}. This may be enough to link them to the, by the
time LISA flies hopefully well determined, merger rates of neutron star and
black hole binaries.

\subsection{Globular clusters}

A special situation occurs in globular clusters: large assemblies of
up to a million old stars in a relatively small volume forming very
dense stellar systems. Binaries play in important role in the
evolution and stability of the clusters \citep[see][]{hmg+92}. From
X-ray observations it has become clear that there is an overabundance
of ultra-compact X-ray binaries in globular clusters, compared to the
disc of the Galaxy. This could in principle be very interesting for
gravitational wave detectors, that are particularly sensitive to
ultra-short periods binaries \citep[e.g.][]{bpr01}. It would be even
more interesting would be if also white dwarfs binaries would be
overabundant in globular clusters. Indeed in some cluster simulations
it has been found that there are more close double white dwarfs in a
dense environment, in particular massive ones that have been proposed
as possible type Ia supernova progenitors
\citep{2002ApJ...571..830S,2003ApJ...589..179H}. However,
\citet{2006MNRAS.372.1043I} do not find such
enhancement. Interestingly, \citet{2008ApJ...679L..29B} find a
puzzling white dwarf sequence in the globular cluster NGC 6791 which
they propose is due to a large fraction (34\%) of double white
dwarfs. So if indeed double white dwarfs are formed more easily in
globular clusters, LISA will see stronger signal from them and thus
may contribute to the study of dynamical formation of binaries in
dense stellar systems.

Another interesting aspect of dynamical interactions in globular
clusters is the possibility of forming eccentric white dwarf binaries,
while in the Galactic disc eccentricity typically is a clear sign of
the presence of a neutron star in the
binary. \citet{2007ApJ...665L..59W} investigate this possibility and
conclude that indeed eccentric binaries are formed and may very well
be detected by LISA. If so, the strong interaction at periastron
passage will give unique information on the internal structure of the
white dwarf and the (tidal) interaction in the system
\citep{2008PhRvL.100d1102W}.

\section{Conclusions and Outlook}

It is clear that there are many interesting developments in
Astrophysics with consequences for LISA, in particular the advent of
large-scale surveys, that will discover a lot of sources that are
interesting for LISA and likely will lead to a much better
understanding of the Galactic population of ultra-compact binaries
before LISA flies. Already now for a number of systems the parameters
are detemined accurately enough that they can serve as varification
binaries. However, LISA is particularly sensitive to many binaries
that are difficult to observe with traditional instruments and will
discover thousands of them. In particular these ultra-short period
binaries are particularly interesting for the physics questions
regarding the (tidal) interaction between compact objects.

Some special attention has been given to globular clusters as these
large assemblies of stars may harbour some very interesting objects
such as eccentrin double white dwarf binaries. It may even be so that
the formation of LISA sources in globular clusters is strongly
enhanced, something that will be tested with the LISA measurements.

\ack It is a pleasure to thank my colleagues for all the interactions
that have been helpful for writing this article. Financial support is
acknowledged from the LKBF and NWO grants VENI 639.041.405 and VIDI
639.042.813

\bibliography{journals,binaries}
\bibliographystyle{iop_harvard}

\end{document}